\documentclass[prd,twocolumn,showpacs,nofootinbib,amsmath,amssymb,floatfix,superscriptaddress,showkeys]{revtex4-2}
\usepackage{multirow}

\usepackage{graphicx}
\usepackage{epstopdf}
\usepackage{dcolumn}
\usepackage{bm}
\usepackage{threeparttable}
\usepackage{subfigure}
\usepackage{color,txfonts}
\usepackage{ulem}
\usepackage{makecell}
\definecolor{blue0}{rgb}{0,0,0.6}
\usepackage[colorlinks,linkcolor=blue0,anchorcolor=blue0,citecolor=blue0,urlcolor=blue0]{hyperref}

\newcommand{\beq}{\begin{equation}}
\newcommand{\eeq}{\end{equation}}
\newcommand{\beqa}{\begin{eqnarray}}
\newcommand{\eeqa}{\end{eqnarray}}

\begin{document}
\title{Investigating the correlations between IceCube high-energy neutrinos and Fermi-LAT $\gamma$-ray observations. II.}

\author{Ming-Xuan Lu}
\affiliation{Laboratory for Relativistic Astrophysics, Department of Physics, Guangxi University, Nanning 530004, China}
\author{Yun-Feng Liang}
\email[]{liangyf@gxu.edu.cn}
\affiliation{Laboratory for Relativistic Astrophysics, Department of Physics, Guangxi University, Nanning 530004, China}
\author{Xuerui Ouyang}
\affiliation{Laboratory for Relativistic Astrophysics, Department of Physics, Guangxi University, Nanning 530004, China}
\author{Rong-Lan Li}
\affiliation{Key Laboratory of Dark Matter and Space Astronomy, Purple Mountain Observatory, Chinese Academy of Sciences, Nanjing 210023, China}
\author{Xiang-Gao Wang}
\email[]{wangxg@gxu.edu.cn}
\affiliation{Laboratory for Relativistic Astrophysics, Department of Physics, Guangxi University, Nanning 530004, China}

\date{\today}
\begin{abstract}
Given that gamma rays with energies larger than TeV are severely absorbed by background radiation fields, for many extragalactic sources, the GeV-TeV gamma-ray observations are the messengers that are closest in energy to the TeV-PeV neutrinos observed by IceCube. Investigating whether there is a correlation between the gamma-ray and neutrino observations can help us identify high-energy neutrino sources and determine which sources are the main contributors to the all-sky diffuse neutrino flux of IceCube. In previous work, we have already studied the possible gamma-neutrino correlations by analyzing 10 years of IceCube public muon-track data. In this work, we further investigate such correlations by employing the IceCube p-value sky map of the scan for point sources. We examine the spatial associations of hotspots in the neutrino sky map with various gamma-ray source samples: the third Fermi-LAT catalog of high-energy sources (3FHL), LAT 14-year source catalog (4FGL), the fourth catalog of active galactic nuclei (4LAC) and subsets of these samples. 
Among all the samples, the 3FHL sample shows a possible correlation with the neutrino hotspots with a pre-trial p-value of $9.0\times10^{-5}$ ($\sim 3.9\,\sigma$), corresponding to a post-trial significance of $\sim1.7\,\sigma$. 
However, this is found to be caused by three already known neutrino sources/source candidates: NGC 1068, TXS 0506+056, and PKS 1424+240. 
\end{abstract}

\maketitle
\section{Introduction}
\label{sec_introduction}

The TeV-PeV diffuse astrophysical neutrino flux detected by IceCube~\cite{Aartsen:2013jdh, IceCube:2014stg, Aartsen:2015knd, IceCube:2016umi, Aartsen:2020aqd, Abbasi:2020jmh, IceCube:2021uhz} opens a new window for astrophysics~\cite{Kistler:2006hp, Beacom:2007yu, Murase:2010cu, Murase:2013ffa, Murase:2013rfa, Ahlers:2014ioa, Tamborra:2014xia, Murase:2014foa, Bechtol:2015uqb, Kistler:2016ask, Bartos:2016wud, Sudoh:2018ana, Bartos:2018jco, Bustamante:2019sdb, Hovatta:2020lor} and particle physics~\cite{Beacom:2002vi,GonzalezGarcia:2005xw,Hooper:2005jp,Ioka:2014kca,Ng:2014pca,Aartsen:2017kpd,Bustamante:2017xuy,Zhou:2019vxt,Zhou:2019frk,Zhou:2021xuh,IceCube:2021rpz,Guo:2023axz,Lu:2024jbq}. However, it is still not clear what astrophysical objects these neutrinos mainly originate from. A large number of works have been carried out to identify the sources of these astrophysical neutrinos \cite{Abbasi:2010rd,Aartsen:2013uuv,Aartsen:2014cva,ANTARES:2015moa,Aartsen:2016oji,IceCube:2017der,Aartsen:2018ywr,Aartsen:2019fau,IceCube:2019lzm,IceCube:2020svz}. The two most promising sources of TeV-PeV neutrinos discovered by IceCube are TXS 0506+056 and NGC 1068~\cite{IceCube:2018dnn,IceCube:2018cha,Aartsen:2019fau,icecube2022evidence}. TXS 0506+056 is found to be temporally and spatially correlated with a $\sim$300 TeV IceCube neutrino event IC-170922A~\cite{IceCube:2018dnn}; furthermore, a neutrino flare from the direction of TXS 0506+056 was detected between 2014 and 2015~\cite{IceCube:2018cha}. In a time-integrated analysis of searching for neutrino point sources using IceCube data between 2011 and 2020, neutrino emission from the Seyfert galaxy NGC 1068~\cite{icecube2022evidence} is found with a global significance of $4.2 \sigma$. More recently, the IceCube Collaboration reported the detection of neutrino emission from the Galactic plane using cascade events with a significance up to 4.7$\sigma$~\cite{icecube2023observation}.

However, the above two point sources and the emission from the Galactic plane contribute only a small fraction of the entire diffuse astrophysical neutrino flux of IceCube, with the origin of the majority of the flux unclear. The high-energy neutrinos detected by IceCube are most likely to be generated by a large number of unresolved extragalactic sources. Many types of astrophysical sources have been considered as high-energy neutrino sources, including gamma-ray bursts~\cite{Waxman:1997ti,Abbasi:2009ig,2012ApJ...752...29H,Aartsen:2014aqy}, star-forming galaxies and starburst galaxies~\cite{Loeb:2006tw,He:2013cqa,Lunardini:2019zcf}, blazars and non-blazar active galactic nuclei (AGNs)~\cite{Stecker:1991vm,Halzen:1997hw,Atoyan:2001ey}, tidal disruption events \cite{Wang:2015mmh}, and some other objects \cite{Pasumarti:2023apw,Bouri:2024ctc}.
Most of the works that search for neutrino emission from these objects display no significant signals, {and therefore they are probably not the only/primary sources of IceCube's diffuse neutrino emission}. For instance, by analyzing the IceCube observations toward the directions of gamma-ray blazars, Refs.
~\cite{IceCube:2016qvd,Hooper:2018wyk,Smith:2020oac,Yuan:2019ucv} found that such a population of sources can contribute at most $\sim$15\% of the diffuse neutrino flux. \citet{Zhou:2021rhl} studied the correlation between the radio-bright AGNs and TeV-PeV IceCube neutrinos, finding no strong correlation between them, implying these radio-bright AGNs contribute no more than 30\% of the all-sky diffuse neutrino emission.

Recently, there have been some works on identifying high-energy neutrino sources displaying positive results. Ref.~\cite{Neronov:2023aks} selected two nearby Seyfert galaxies according to the intrinsic luminosity in the X-ray energy band and searched for neutrino emission from the two sources using 10 years of IceCube public muon-track data. Evidence of neutrino signals from both sources is found which supports that Seyfert galaxies are one population of TeV-PeV neutrino emitters. Refs.~\cite{2021NatAs...5..510S,2022PhRvL.128v1101R,vanVelzen:2021zsm}  performed ZTF follow-up observations of IceCube neutrino alert events and found that the tidal disruption events (TDEs) AT2019dsg, AT2019fdr, and AT2019aalc are temporally and spatially correlated with IC191001A, IC200530A, and IC191119A, respectively, implying that TDEs are likely to be an important class of neutrino sources. Note however that \citet{Liao:2022csg} suggests the event IC-191001A is possibly related to the blazar GB6 J2113+1121 instead of AT2019dsg. 
By comparing the spatial positions of blazars with those of the IceCube alert events or the hotspots in the p-value skymaps of neutrino point source scan, several works \cite{Plavin:2022oyy,Buson:2022fyf,Buson:2023irp} claim that there exists high confidence evidence proving that blazars are a population of extragalactic neutrino sources. However, a subsequent study argued that the strong correlation could be due to a statistical fluctuation and possibly the spatial and flux nonuniformities in the blazar sample \cite{Bellenghi:2023yza}.

This work will investigate the correlation between IceCube neutrinos and Fermi-LAT gamma-ray sources. One motivation is that high-energy astrophysical neutrinos are expected to be generated in a hadronic process, so that gamma rays will always be produced simultaneously accompanying the neutrino production. Since TeV-PeV gamma rays are severely absorbed by background radiation fields, for most extragalactic sources, the GeV-TeV gamma-ray observations from Fermi-LAT are the messengers with energies closest to the TeV-PeV neutrinos observed by IceCube. {It should be noted that the correlation between neutrinos and gamma rays may not be straightforward. According to the status of art of hadronic models, when neutrinos are produced, co-spatially produced gamma rays have a high probability of being absorbed within the emission region, and the gamma rays will be reprocessed and appear at lower energies (from X-rays to MeV) \citep{Murase:2020lnu}. However, considering that gamma-ray emission is a good indicator of a source’s capability of particle acceleration, studying the correlation between neutrino data and gamma-ray observations is still helpful for identifying potential neutrino sources \citep{Murase:2020lnu}.}
In our previous work, we have investigated the correlations between various Fermi-LAT gamma-ray samples and IceCube neutrinos \cite{Li:2022vsb}. Some other analyses on Fermi-LAT catalogs using IceCube data include Refs.\cite{IceCube:2016qvd,Hooper:2018wyk,Smith:2020oac,Huber:2019lrm,Luo:2020dxa,IceCube:2021slf,IceCube:2022zbd,IceCube:2021xar}. This work will continue to test whether there exists a correlation between the GeV-TeV observations of Fermi-LAT and the IceCube neutrino observations. The difference from the previous work, \citet{Li:2022vsb} (hereafter L22), is that L22 handles directly the 10-year IceCube muon-track data using a likelihood ratio method and search for excess signals from the directions of Fermi-LAT gamma-ray point sources. In this work, we adopt an alternative approach as in Refs.~\cite{Buson:2022fyf,Buson:2023irp} (hotspots analysis), which claim the finding of evidence of correlation by comparing the spatial positions of the 5th edition of the Roma-BZCAT Multifrequency Catalogue (5BZCAT) of Blazars with the hotspots in the IceCube p-value sky map. The sky map is obtained from the northern-sky blind searches of neutrino point sources \cite{icecube2022evidence}. 

Advantages of the hotspots analysis compared to the direct likelihood analysis of 10 years of muon-track data (likelihood analysis) include: at present only the muon-track data in the 2008-2018 period is publicly released, while the p-value sky map from Ref.~\cite{icecube2022evidence} is based on the data in the period 2011-2022, the hotspots analysis allows for the utilization of longer IceCube observations with higher sensitivity. In addition, if there exist systematic biases in the neutrino source localization that cause the neutrino signal to deviate from the true source position, a direct likelihood search at the source position would yield a null result even if the signal exceeds the background, whereas the hotspots analysis could take this into account by introducing an association radius. The hotspots analysis also does not require a priori assumptions of weighting factors. 

We will search for evidence of correlations between various Fermi-LAT gamma-ray samples and IceCube neutrino hotspots. The gamma-ray catalogs/samples considered in this work include: the third Fermi-LAT catalog of high-energy sources (3FHL), LAT 14-year source catalog (4FGL), the fourth catalog of active galactic nuclei (4LAC) and subsets of these catalogs.

\begin{table*}[!htbp]
	\centering
	\caption{Fermi-LAT samples considered in this work.}
	\begin{tabular}{p{3cm}p{2.5cm}p{1.5cm}p{2cm}p{1.5cm}p{1.5cm}p{1.0cm}}
		\hline
		\hline
		Catalog Name & Energy Range$^a$ & $N_{\rm src}^b$ & $~~P^{~c}$ & $N_{\rm assoc}^d$ & $R_{\rm assoc}^{~e}$ & $L_{\rm min}^e$ \\
		\hline
		4FGL-DR4 & \multirow{4}{*}{50 MeV-1 TeV} & 5056 & 5.60 $\times$ 10$^{-4}$  & 47 & 1.3 & 3.0  \\
		4FGL-DR4 Blazars &  & 3515 & 1.78 $\times$ 10$^{-3}$ & 37 & 1.2 & 3.0  \\
		4FGL-DR4 BL Lacs &  & 1407 &  1.03 $\times$ 10$^{-3}$ & 3 & 0.2 & 3.5  \\
		4FGL-DR4 FSRQs &  & 783 & 3.04 $\times$ 10$^{-1}$ & 4 & 1.15 & 3.5   \\
		\hline
		4LAC-DR3 & \multirow{4}{*}{50 MeV-1 TeV} & 3407 &  1.06 $\times$ 10$^{-3}$ & 37 & 1.2 & 3.0 \\
		4LAC-DR3 Blazars &  & 3342 &  8.40 $\times$ 10$^{-4}$ & 40 & 1.3 & 3.0 \\
		4LAC-DR3 BL Lacs &  & 1379 & 7.90 $\times$ 10$^{-4}$ & 3 & 0.2 & 3.5 \\
		4LAC-DR3 FSRQs &  & 755 & 3.05 $\times$ 10$^{-1}$ & 1 & 0.6 & 4.0\\
		\hline
            3FHL & \multirow{4}{*}{10 GeV-2 TeV} & 1219 & 9.00 $\times$ 10$^{-5}$ & 3 & 0.25 & 4.0 \\
            3FHL Blazars &  & 1080 & 1.42 $\times$ 10$^{-3}$ & 2 & 0.2 & 4.0  \\
            3FHL BL Lacs &  & 697 & 7.40 $\times$ 10$^{-4}$ & 2 & 0.2 & 4.0  \\
            3FHL FSRQs &  & 165 & 1.20 $\times$ 10$^{-1}$ & 2 & 1.15 & 3.5  \\
            \hline
            5BZCAT$^f$ & - & 3461 & 1.12 $\times$ 10$^{-3}$ & 42 & 1.25 & 3.0  \\
		\hline
		\hline
	\end{tabular}
\begin{tablenotes}
\item $^a$ The energy range of the Fermi-LAT data used to construct the catalog. \\
\item $^b$ Number of sources included in the sample. The sources categorized as 'PSR' or in the $|b|<10^{\circ}$ region have been excluded.
\\
\item $^c$ The minimum p-value in the correlation analysis.\\
\item $^d$ The number of associations between Fermi-LAT/BZcat catalogs and neutrino hotspots corresponding to the minimum p-value.
\item $^e$ The $R_{\rm assoc}^{~e}$ and $L_{\rm min}^e$ values that lead to the minimum p-value.
\item $^f$ This is not a Fermi-LAT sample; we include it here for comparison. Also note that since 5BZcat is a compiled catalog with pronounced non-uniform sky distribution of sources, the $p$-value result here may involve applicability issue of the randomization method.
See Sec.~\ref{sec:bzcat} for related discussions.

\end{tablenotes}
	\label{tab:tab1}
\end{table*}

\begin{figure*}[t]
\centering
\includegraphics[width=0.4\textwidth]{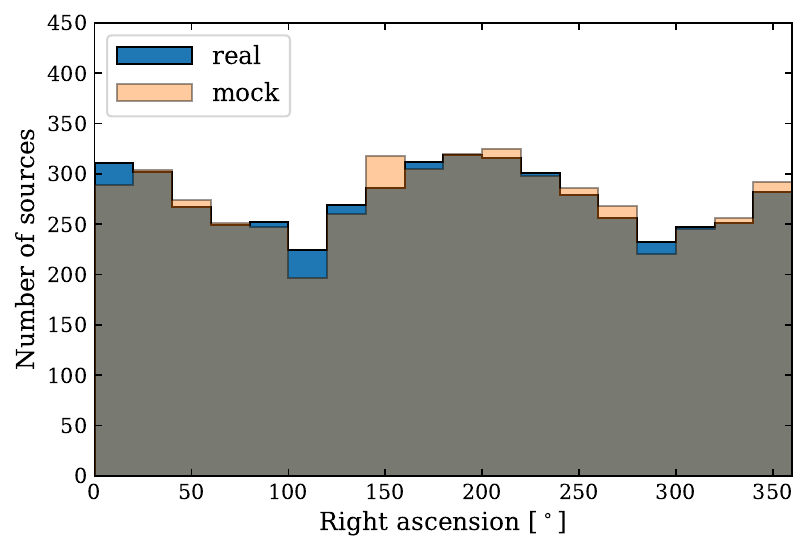}
\includegraphics[width=0.4\textwidth]{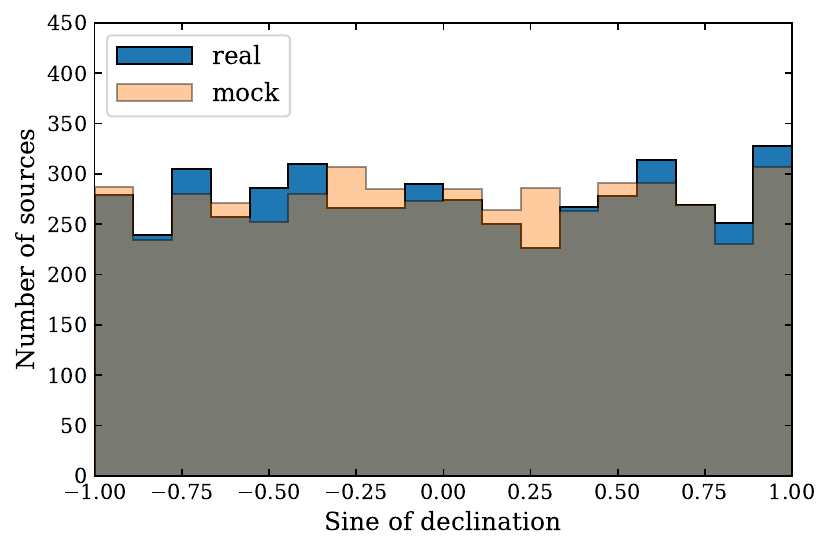}
\caption{A comparison of the right ascension (RA) and declination (Dec) distributions between the real 4FGL catalog and one realization of the mock catalogs. The distributions are well consistent, implying the generated mock catalog has the same spatial distribution with the real one.}
\label{fig:distr}
\end{figure*}

\section{Fermi-LAT gamma-ray samples}
\label{sec_data}
The Large Area Telescope (LAT) on board the Fermi satellite is a wide field-of-view (FOV) imaging gamma-ray telescope, which detects gamma-ray photons in the energy range from $\sim30\,{\rm MeV}$ to $>300\,{\rm GeV}$ \cite{Fermi-LAT:2009ihh}.
Since 2008, the Fermi-LAT continuously surveys the entire sky. Its observation period and FOV overlap with the IceCube observation considerably. With more than 16 years of observations, a variety of source catalogs of Fermi-LAT have been compiled and released. To investigate the correlations between GeV-TeV gamma-ray sources and TeV-PeV neutrinos, we consider the following Fermi-LAT samples.

Among all the Fermi-LAT catalogs, the sources contained in the Third Catalog of Hard Fermi-LAT Sources (3FHL sample)\footnote{\url{https://fermi.gsfc.nasa.gov/ssc/data/access/lat/3FHL/}}~\cite{Fermi-LAT:2017sxy} represent the population of GeV gamma-ray sources has hardest spectra, which are therefore more likely to be accelerators of TeV-PeV particles. The 3FHL is constructed based on 7 years of Fermi-LAT data in the 10 GeV-2 TeV energy range~\cite{Fermi-LAT:2017sxy}. 
Besides the 3FHL, other samples considered in this work include, the gamma-ray sources of the fourth Fermi-LAT catalog (4FGL-DR4, for Data Release 4)\footnote{\url{https://fermi.gsfc.nasa.gov/ssc/data/access/lat/14yr_catalog/}}~\cite{Fermi-LAT:2019yla,Fermi-LAT:2022byn} and the fourth catalog of active galatic nuclei (4LAC-DR3, for Data Release 3)\footnote{\url{https://fermi.gsfc.nasa.gov/ssc/data/access/lat/4LACDR3/}}~\cite{Fermi-LAT:2019pir}. The adopted source catalog files are as follows: {\tt gll$\_$psc$\_$v34.fit} for 4FGL, {\tt table-4LAC-DR3-h.fits} for 4LAC, and {\tt gll$\_$psch$\_$v13.fit} for 3FHL.
We also consider some subsamples. A blazar subsample contains all sources classified as blazars in the catalog. Blazars are further divided into two subclasses: flat spectrum radio quasars (FSRQs) and BL Lacertae objects (BL Lacs). Totally, there are three subsamples (Blazar, BL Lac, FSRQ).

All these samples are summarized in Table~\ref{tab:tab1}. For each sample, we exclude the sources categorized as 'PSR' (pulsar) from the catalog. This is because pulsars primarily emit gamma rays via leptonic mechanisms (curvature radiation of relativistic electrons), which do not produce neutrinos. However, millisecond pulsars (sources categorized as 'MSP') are retained in the sample, as MSPs are frequently associated with binary systems and globular clusters. We have tested that whether or not to exclude PSRs would only have a very small effect on the results since no hotspot is associated with these PSRs. To avoid the complexity of the Galactic Plane, we only select $\left | b \right |\geqslant10^{\circ}$ sources from the catalogs.

\section{Analysis Method}
\label{sec_ana}

The IceCube Neutrino Observatory detects neutrinos by detecting the Cherenkov light emitted by relativistic secondary charged particles from neutrino interactions~\cite{Achterberg:2006md}.
In Ref.~\cite{icecube2022evidence}, IceCube Collaboration performed a blind search for high-energy neutrino point sources in the northern sky using the IceCube muon-track data from 2011 to 2020. They binned the entire sky into 786432 pixels (i.e. healpix pixelization with {\tt nside=256}), and placed a putative neutrino point source at the center of each pixel for all pixels with $-3^\circ\leqslant\delta<81^\circ$. A likelihood analysis is performed to derive the statistical significance of existing excess neutrino emission at the corresponding position. 
A p-value sky map is obtained from this search.
The local significance of the neutrino excess is represented by a $-\log_{10} P$ value, where $P$ is the p-value, i.e. the probability of obtaining the observed result due to a background fluctuation. We follow \citet{Plavin:2022oyy} and \citet{Buson:2022fyf} to denote the negative logarithm of local p-values mapping the neutrino sky as $L = -\log_{10}P$ to avoid confusion with the p-values in the later correlation analyses. 
A larger $L$ value means a more significant neutrino excess beyond the background at that direction and thus a higher probability of the existence of a real neutrino source. So the points with large $L$ can be regarded as neutrino point source candidates.

In this work, we adopt this p-value sky map of Ref.~\cite{icecube2022evidence} which is based on the 2011-2020 data and covers the northern sky of $-3^\circ\leqslant\delta<81^\circ$. We first define the hotspots in the sky map to be used for subsequent analysis. For better comparison with previous results, we use the same criteria for defining hotspots as in Ref.~\cite{Buson:2023irp}. The hotspots in the p-value skymap are defined as the pixels with large $L$ values. We pick out the pixels in the sky map with $L$ (i.e. $-\log_{10} P$) larger than a certain predefined threshold ($L_{\rm min}$). Adjacent pixels within a small region will be treated as one hotspot, namely two independent hotspots are required to have an angular distance of $>1.5^\circ$ ($1.5^\circ$ is the median angular uncertainty of the muon-track data of IceCube \cite{icecube2022evidence,Buson:2023irp}). Different choices of $L_{\rm min}$ lead to varying numbers of hotspots. We consider three $L_{\rm min}$ values: 3.0, 3.5, and 4.0, which result in 81\footnote{ \citet{Buson:2023irp} reports 82 total hotspots with $L>3.0$ while we here reports 81. Please see Appendix~\ref{app:hotspot} for the reason. Since the removed hotspot lies within the Galactic plane with $|b|<10^\circ$, whether or not it is included does not affect the results of the subsequent analysis.}, 34, and 17 hotspots, respectively. All 81 hotspots with $L>3.0$ are listed in Appendix \ref{app:hotspot}. The same as Ref.~\cite{Buson:2023irp}, we only consider the hotspots with $|b|\geqslant10^\circ$. After the Galactic plane cut, 66, 29 and 13 hotspots remain for $L_{\rm min}=3.0$, $3.5$, and $4.0$, well consistent with the hotspot numbers in Ref.~\cite{Buson:2023irp}.

\begin{figure*}[htbp]
\centering
\subfigure[~TOTAL samples]{
\includegraphics[width=0.45\textwidth]{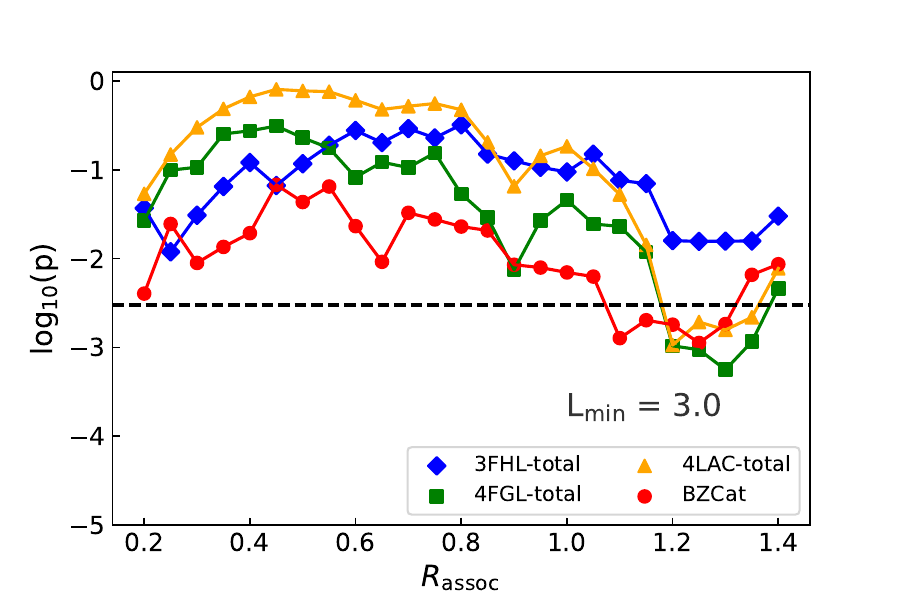}
}
\subfigure[~Blazar subsamples]{
\includegraphics[width=0.45\textwidth]{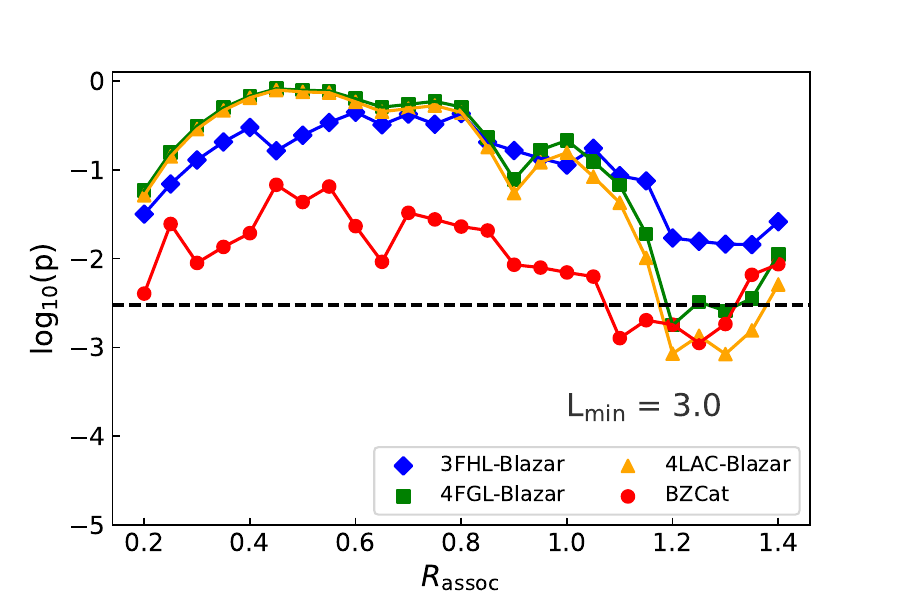}
}\vspace{-1em}
\subfigure[~BL Lac subsamples]{
\includegraphics[width=0.45\textwidth]{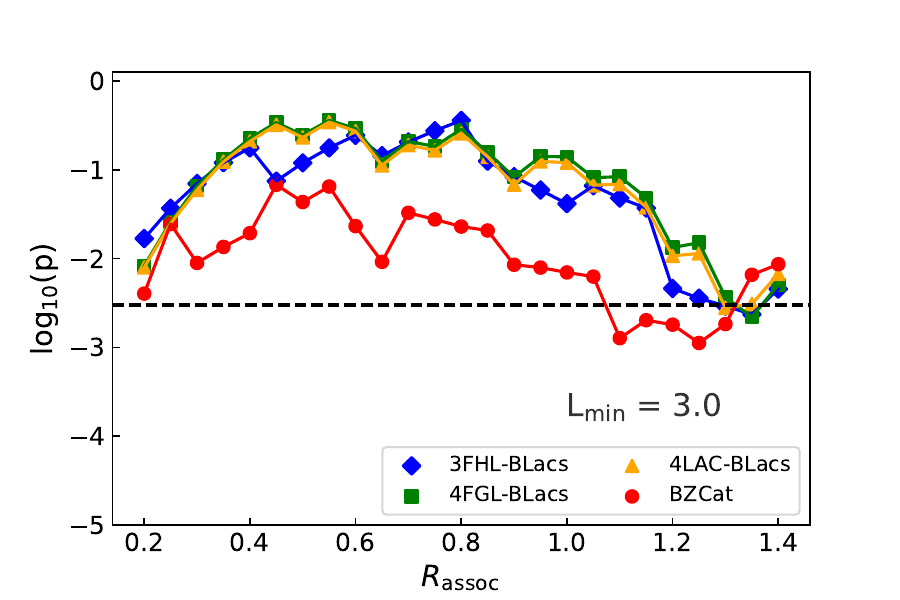}
}
\subfigure[~FSRQ subsamples]{
\includegraphics[width=0.45\textwidth]{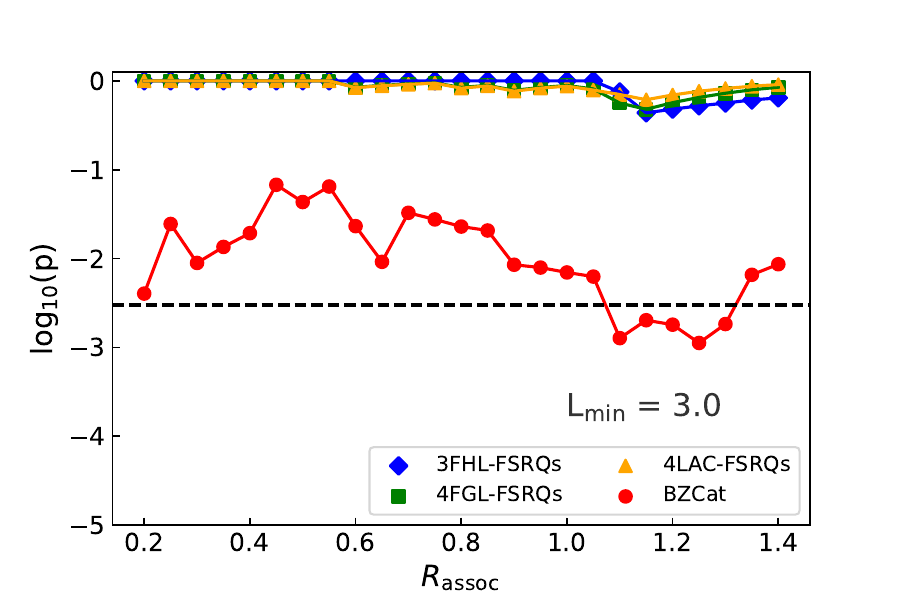}
}
\caption{Pre-trial p-values for the correlations between various Fermi-LAT gamma-ray samples (diamond, triangle, and square points for 3FHL, 4LAC, and 4FGL samples, respectively) and IceCube neutrino hotspots in the northern sky, as a function of the association radius ($R_{\rm assoc}$) and for a minimum significance threshold for the hotspots of $L_{\rm min}=3$. The horizontal dashed line shows the significance level of $3\sigma$. As a reference, the result of the correlation analysis with the BZCat (see Sec.~\ref{sec:bzcat}) is also shown (circle points), which is the same in the 4 panels.}
\label{fig:L3.0}
\end{figure*}

\begin{figure*}[htbp]
\centering
\subfigure[~TOTAL samples]{
\includegraphics[width=0.45\textwidth]{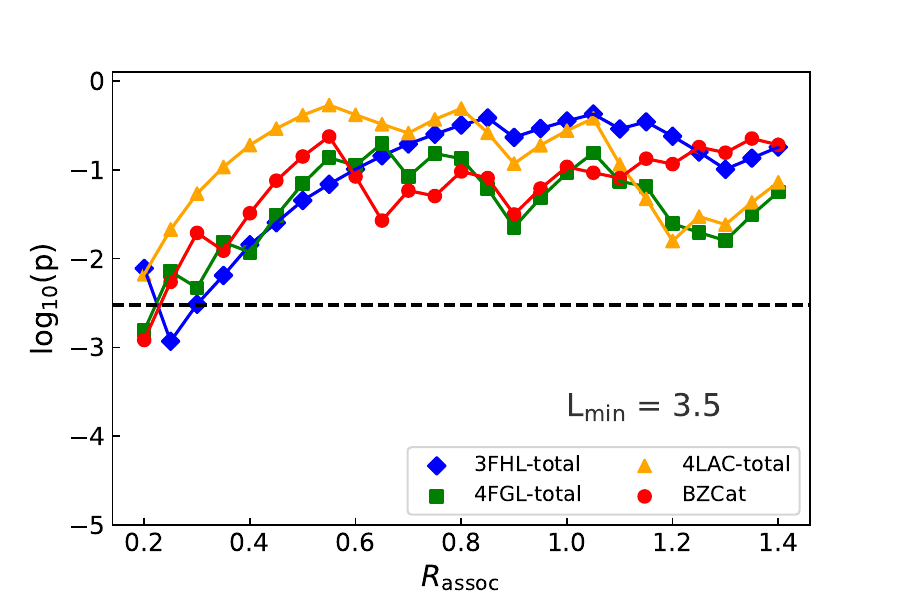}
}
\subfigure[~Blazar subsamples]{
\includegraphics[width=0.45\textwidth]{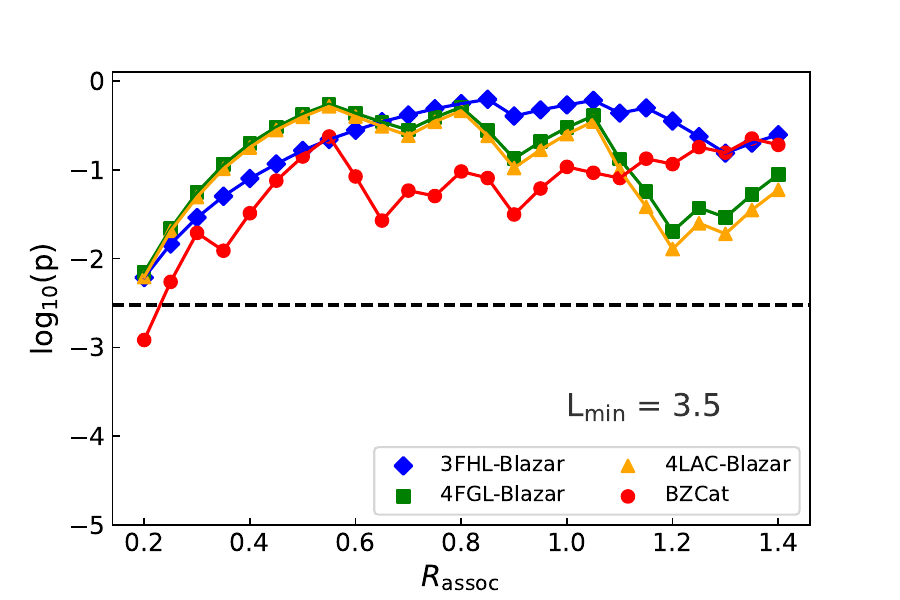}
}\vspace{-1em}
\subfigure[~BL Lac subsamples]{
\includegraphics[width=0.45\textwidth]{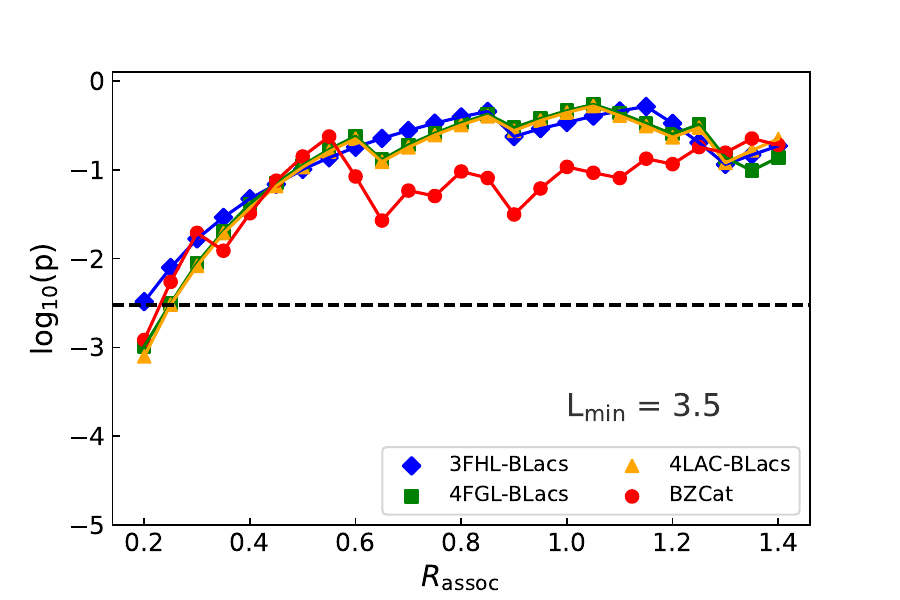}
}
\subfigure[~FSRQ subsamples]{
\includegraphics[width=0.45\textwidth]{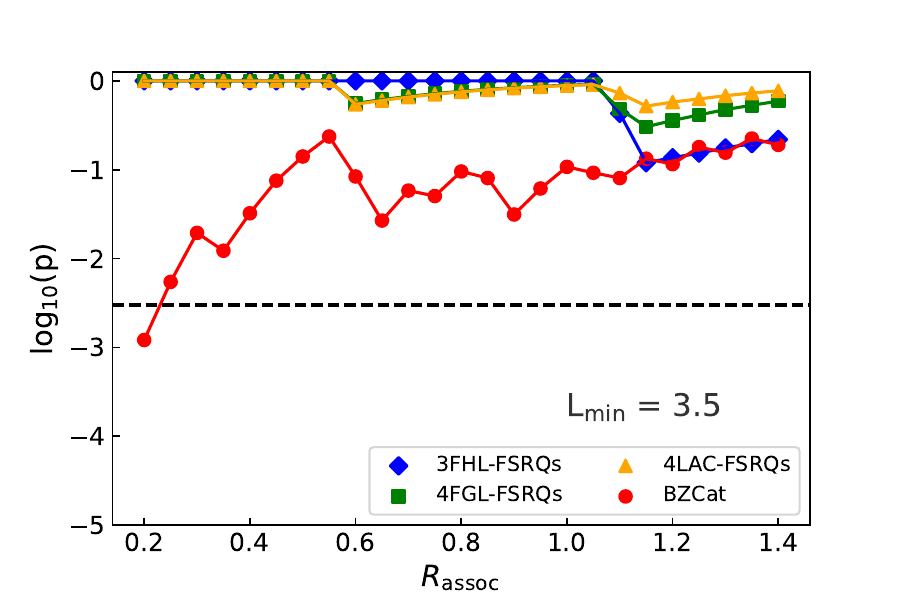}
}
\caption{The same as Fig.~\ref{fig:L3.0} but for $L_{\rm {min}} = 3.5$.}
\label{fig:L3.5}
\end{figure*}

\begin{figure*}[htbp]
\centering
\subfigure[~TOTAL samples]{
\includegraphics[width=0.45\textwidth]{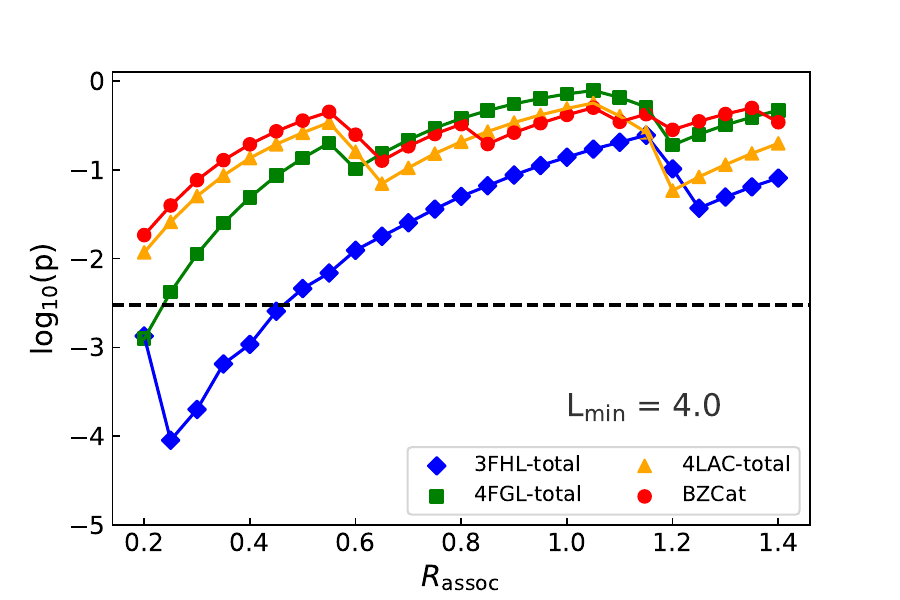}
}
\subfigure[~Blazar subsamples]{
\includegraphics[width=0.45\textwidth]{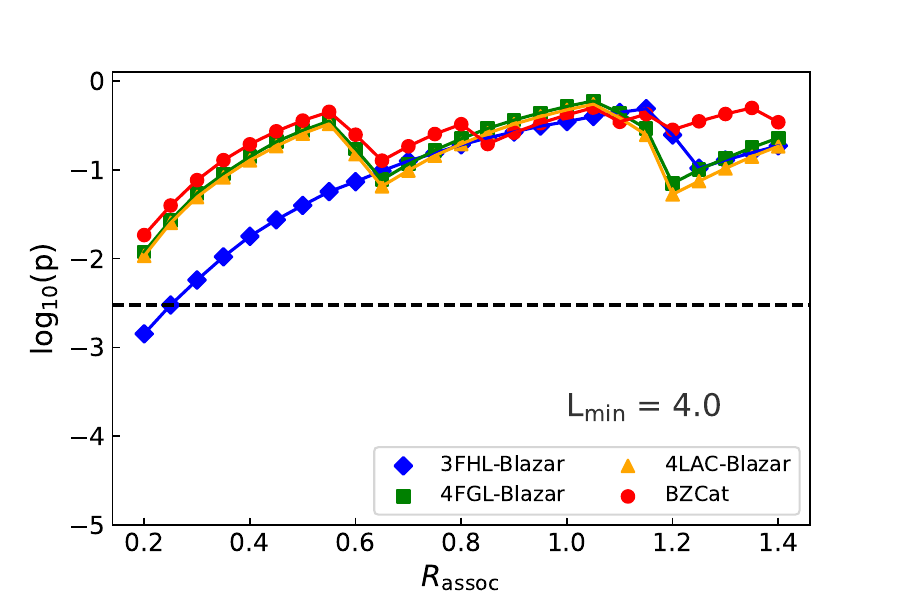}
}\vspace{-1em}
\subfigure[~BL Lac subsamples]{
\includegraphics[width=0.45\textwidth]{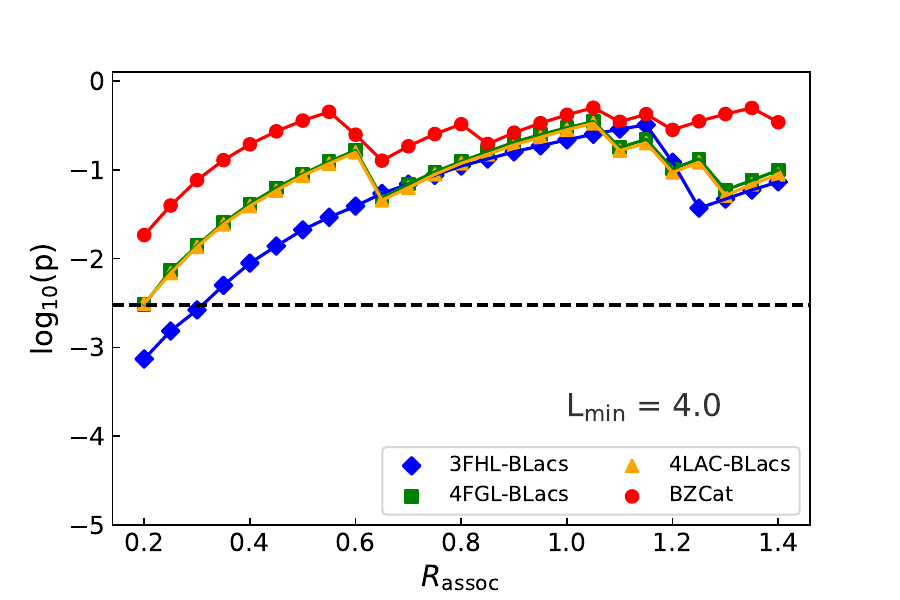}
}
\subfigure[~FSRQ subsamples]{
\includegraphics[width=0.45\textwidth]{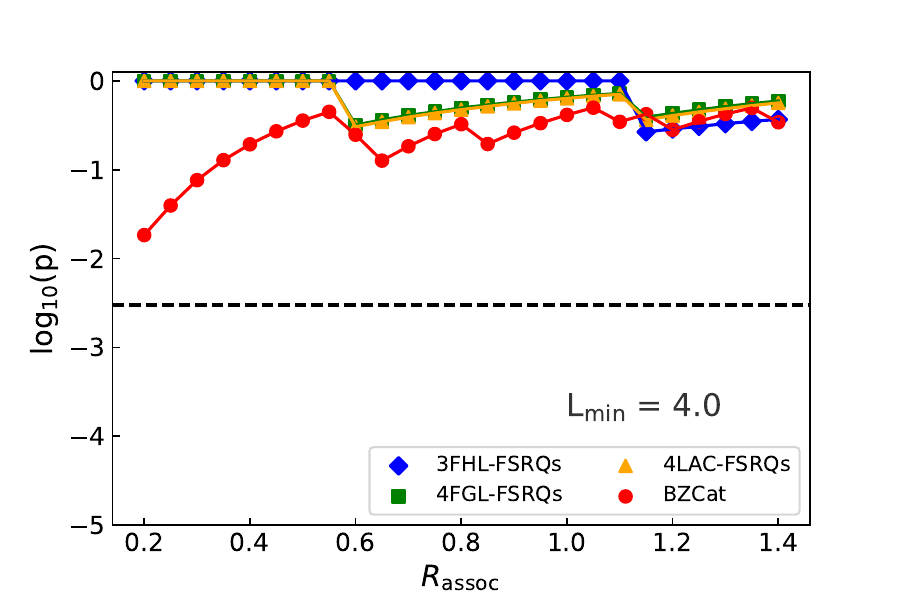}
}
\caption{The same as Fig.~\ref{fig:L3.0} but for $L_{\rm {min}} = 4.0$.}
\label{fig:L4.0}
\end{figure*}

If the angular distance between a source and its nearest hotspot is less than a certain radius $R_{\rm assoc}$, we treat this source and the hotspot as positional associated. To be conservative, one neutrino hotspot will be matched with only one source at most. We scan all sources in the catalog and count the number of associations. We only consider the sources with $|b|\geqslant10^\circ$ and $-3^\circ\leqslant\delta<81^\circ$ where $\delta$ is the declination. To know whether an association number $N$ can be interpreted as a chance coincidence or it implies a high confidence correlation between the catalog and the neutrino hotspots, we carry out Monte Carlo simulation to give the p-value corresponding to the association number $N$. In the simulation, we keep the positions of the hotspots unchanged and randomly generate pseudo sources. For each simulation, the number of pseudo sources is the same as that of the real catalog, which is listed in the third column of Table \ref{tab:tab1}.

Considering that the Fermi-LAT gamma-ray sources are more symmetrically distributed in Galactic coordinates, we generate the longitudes and latitudes for mock sources in Galactic coordinates. To guarantee the generated mock source list has the same $l$ and $b$ distributions as those of the real catalog, we each time directly sample an $l'$ from the real $l$ list and a $b'$ from the real $b$ list. We obtain a $(l',b')$ pair by this way.
Further, 
we impose an additional random shift to $(l',b')$ within a circle region of $5^\circ$ radius.
We choose the $5^\circ$ randomization radius so that our sampling can basically cover the entire sky for the samples with $N_{\rm src}\gtrsim700$ (except for the 3FHL FSRQ sample) while maintaining the overall distribution features of the real sources. We have tested that the choice of other randomization radii does not affect the main results/conclusions of this work (please see Fig.~\ref{fig:different-deg} in Appendix~\ref{app:cross-check} for the tests).
The final coordinates of the mock source are $\hat{\bm{p}}+\Delta\hat{\bm{p}}$, where $\hat{\bm{p}}=(l',b')$ and $|\Delta\hat{\bm{p}}|<5^\circ$. Meanwhile, we also use KS tests to further ensure that the mock sources preserve the distribution pattern of the actual sources. We apply the KS tests on the RA and Dec parameters. A mock sample with a probability of rejecting the null hypothesis greater than 0.95 is discarded, {where the null hypotheses is that the mock sample and the observations come from the same distribution.} The mock source catalog generated in this way has almost identical spatial distribution as the real catalog (see Fig.~\ref{fig:distr} for a demonstration). Then we apply the $-3^\circ\leqslant\delta<81^\circ$ cut, and perform absolutely the same correlation analysis with the hotspots. 
{It should be noted that in our simulation process, we generate the identical number of all-sky mock sources and then apply the $-3^\circ\leqslant\delta<81^\circ$ cut. In this step, we do not impose the limitation requiring the number of sources within the $-3^\circ$ to $81^\circ$ range to be exactly the same as in the real catalogs. We have tested that the impact on the results is small.}
For each simulation, we obtain an association number $\tilde{N}$. We run $10^5$ simulations, and the p-value corresponding to the $N$ associations of the real catalog can be derived based on the distribution of $\tilde{N}$.

\section{Results}
\label{sec_results}

Utilizing the analysis method described above, we investigate the correlations between various Fermi-LAT gamma-ray source catalogs and IceCube neutrino hotspots. We consider 4FGL, 4LAC and 3FHL catalogs and some subsamples of them. A blazar subsample contains all sources classified as blazars in the catalog. We also further divide the blazars into two subclasses: FSRQs and BL Lacs. Totally, we perform the analyses for three subsamples (Blazar, BL Lac, FSRQ). The results of our analyses are presented in Fig.\ref{fig:L3.0}, Fig.\ref{fig:L3.5} and Fig.\ref{fig:L4.0}, in which we show the local p-value, $p_{\rm local}$, for different choices of $L_{\rm min}$ and $R_{\rm assoc}$. The minimum p-value for each sample and the corresponding association number are listed in Table~\ref{tab:tab1}.

We find that most samples except the FSRQ ones yield minimum p-values of $<3\times10^{-3}$. Among all the analyses, the 3FHL catalog shows the most significant correlation with the neutrino hotspots, $p=9.0\times10^{-5}$, corresponding to a pre-trial significance of $\sim3.9\,\sigma$. 
Besides this, the second lowest p-value of $5.6\times10^{-4}$ comes from the analysis of the 4FGL TOTAL sample.

In this type of analysis, the randomization approaches used to generate mock sources are crucial for the results, and the use of different approaches will likely affect the results obtained (see also Sec.~\ref{sec:bzcat} and Fig.~\ref{fig:5bzcat}). Ideally, a randomization approach needs to be able to maintain both the local and global distributions of the real sources. However, in practice there is a trade-off between ensuring the local pattern is not destroyed while also being as uncorrelated as possible with the actual catalog. In our work, in order to make the mock sources sufficiently uncorrelated, we draw $l$ and $b$ independently from the real $l$ and $b$ lists. This procedure ignores possible correlations between $l$ and $b$, which may destroy the local pattern to some extent.
For this reason we also use other randomization methods to cross-check our results, which are presented in Appendix \ref{app:cross-check}. It can be seen that the results obtained by the different sampling methods have only small differences in the quantitative $\log_{10}(p)$ values, which do not affect the main results of our paper.

Because we change/attempt different $L_{\rm min}$ and $R_{\rm assoc}$ in the analysis of each catalog, implying that we have introduced multiple trials, this effect should be considered and corrected to convert the local p-value to a global post-trial p-value.
The post-trial p-value can be obtained by
\begin{equation}
    p_{\rm global}=1-(1-p_{\rm local})^k,
\end{equation}
where $k$ is the number of independent trials.
We have considered 3 different $L_{\rm min}$ and 25 different $R_{\rm assoc}$, so $k=3\times25=75$.
Further considering we have tested 12 samples in this work, the trials increase to $k=75\times12=900$.
After the trial correction, the above p-values of $9.0\times10^{-5}$ (for 3FHL TOTAL) and $5.6\times10^{-4}$ (for 4FGL TOTAL) correspond to post-trial significances of 1.7 and 1.0 $\sigma$, respectively. The post-trial significance of other samples will be even lower.

However, we think that the trial number of $k=900$ might have been overestimated. One reason is that, the different source catalogs/samples have many overlapped sources, meaning they are not completely independent of each other. {We also test another trial-correction method (harmonic mean p-value method) as adopted in \citet{Kouch:2024xtd}, which derives the post-trial p-value based on a series of p-values obtained from multiple {\it overlapping} blazar subsamples. By adopting this trial-correction method, a significance of $\sim 2.4\sigma$ is obtained for the correlation between the Fermi-LAT catalogs and the hotspots, based on the total of 900 p-values shown in Fig.~\ref{fig:L3.0} $-$ Fig.~\ref{fig:L4.0}. The conclusion remains unchanged, that is, the correlation between the gamma-ray sources and the neutrino hotspots is still not significant.}

We next examine which sources in the 3FHL sample are located close to neutrino hotspots, and lead to the relatively small p-value of $9.0\times10^{-5}$.
In Table~\ref{tab:3src} we list the three sources in 3FHL that are located close to the hotspots.
We can see that, all three sources are those ones that have already been reported to be neutrino sources/source candidates in previous works \cite{IceCube:2018dnn,IceCube:2018cha,Aartsen:2019fau,IceCube:2021xar,icecube2022evidence}.
Therefore, we do not identify any new neutrino source candidate.

\begin{table*}[!htbp]
	\centering
	\caption{Sources in 3FHL associated with neutrino hotspots for $R_{\rm {assoc}} = 0.25^\circ$ and $L_{\rm min}=4.0$.}
	\begin{tabular}{p{3cm}p{1.5cm}p{1.6cm}p{2.5cm}p{2cm}p{2cm}p{1cm}p{1.2cm}}
		\hline
		\hline
		3FHL name & RA$_{\rm {3FHL}}[^{\circ}]$ & DEC$_{\rm {3FHL}}[^{\circ}]$ & Counterpart & \text{RA$_{\rm {hotspot}}[^{\circ}]$} & \text{DEC$_{\rm {hotspot}}[^{\circ}]$} & $L_{\rm {hotspot}}$ & $\Delta\,[^{\circ}]^{\,a}$ \\
		\hline
		3FHL J1427.0+2348 & 216.76 & 23.80 & PKS 1424+240 & 216.91 & 23.81 & 4.18 & 0.143 \\
		3FHL J0509.4+0542 & 77.36 & 5.71 & TXS 0506+056 & 77.34 & 5.53 & 4.13 & 0.178 \\
            3FHL J0242.7-0.002 & 40.68 & $-0.04$ & NGC 1068 & 40.78 & 0.15 & 6.75 & 0.219 \\
		\hline
		\hline
	\end{tabular}
 \begin{tablenotes}
 \item $^a$ The angular separation between the 3FHL position and the position of the hotspot.
 \end{tablenotes}
 \label{tab:3src}
 \end{table*}

\section{Discussion}
\subsection{Constraints on the contribution to IceCube’s diffuse neutrino flux}

If we know the discovery potential of 2011-2020 data for point-source/hotspot detection, by assuming a relation between the neutrino flux $\phi_\nu$ and the gamma-ray flux $f_\gamma$ or the source distance $D_L$ (namely the weighting schemes in the previous works\footnote{The relations/weighting schemes usually considered include: equal weighting, $\gamma$/X-band flux weighting ($\phi_\nu\propto f_\gamma$) and geometrical weighting ($\phi_\nu\propto1/D_L^2$), but our analysis here does not apply to the equal weighting scheme.} \cite{IceCube:2016qvd,Smith:2020oac, Zhou:2021rhl,Li:2022vsb}), for any catalog we can also derive its maximum contribution to the diffuse neutrino flux based on the analysis in this work. We note that for all 3 catalogs (3FHL, 4LAC, 4FGL), there is only 1 source (i.e. NGC 1068) being detected with significance greater than $5\sigma$ \cite{icecube2022evidence}.  
Then, at the 95\% confidence level, the number of sources with expected neutrino fluxes exceeding the $5\sigma$ discovery potential curve must be fewer than 8. This can be known through the binomial distribution, 
\begin{equation}
P(X = k) = \binom{n}{k}\,p^k\,(1-p)^{n-k}
\end{equation}
where $k=1$ is the number of sources actually detected, $n$ is the number of sources expected to exceed the $5\sigma$ discovery potential, and $p=0.5$ is the detection probability for a source with its flux exactly at the discovery potential. 
Here we adopt the $5\sigma$ discovery potential line in Ref.~\cite{icecube2022evidence} (see Figure~S11 of the Supplementary Materials), which is defined as the flux necessary to make a $5\sigma$ discovery with 50\% probability (i.e., a source at this flux has a 50\% probability to be discovered at a local significance of $>5\sigma$).
By setting $P(X\leq1)<0.05$ (because we require a 95\% confidence level), we obtain $n=8$.
Therefore, the number of sources expected to exceed the $5\sigma$ discovery potential curve should not be greater than 8, otherwise the number of sources detected with $>5\sigma$ significance would be more than 1 at the 95\% confidence level, contradicting the actual observation.

By assuming the neutrino flux is proportional to a certain quantity $X$ (e.g., gamma-ray flux $f_\gamma$ or source distance $1/D_L^2$) and requiring that the expected neutrino fluxes exceed the $5\sigma$ discovery potential curve for exactly 8 sources, we can then determine the proportionality between the neutrino flux $\phi_\nu$ and $X$. 
More specifically, we assume $\phi^{\rm 1\,TeV}_\nu=A\times X$ with $X$ being $f_\gamma$ or $1/D_L^2$ and $\phi^{\rm 1\,TeV}_\nu$ being the neutrino flux at 1 TeV (since the discovery potential is reported at 1 TeV), then we adjust $A$ to make just only 8 sources have $\phi^{\rm 1\,TeV}_\nu$ above the $5\sigma$ discovery line in the Northern sky. 
The upper limit on the flux from the whole catalog can be estimated by summing the fluxes of all sources in the catalog together, $dN_\nu/dE_\nu(E_\nu)=\sum_i \phi^{\rm 1\,TeV}_{\nu,i}/4\pi\times(E_\nu/{\rm 1\,TeV})^{-\gamma}$.
Here, we have assumed the neutrino sources have an averaged intrinsic spectrum of $E_\nu^{-\gamma}$. We set $\gamma=2.0$ or $3.2$, since only the discovery potential lines for these two indices are available in Ref.~\cite{icecube2022evidence}.

\begin{figure*}[htbp]
\centering
\includegraphics[width=0.45\textwidth]{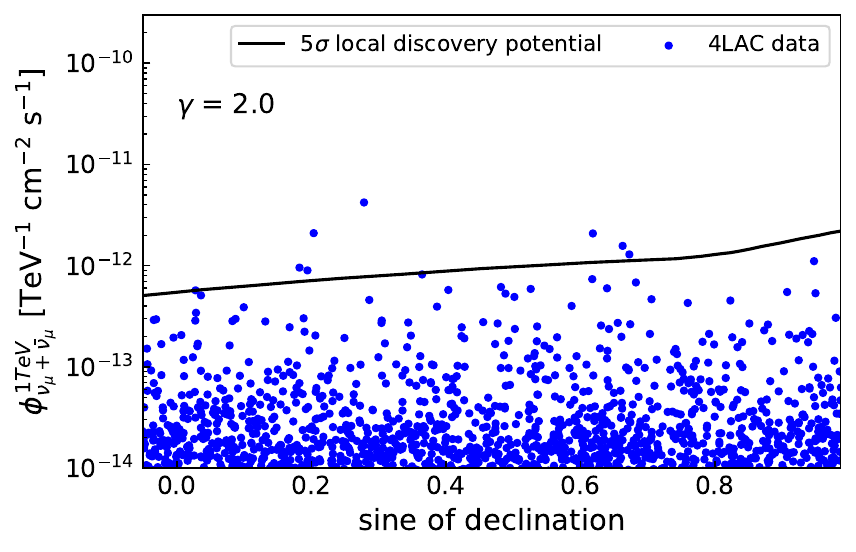}
\includegraphics[width=0.45\textwidth]{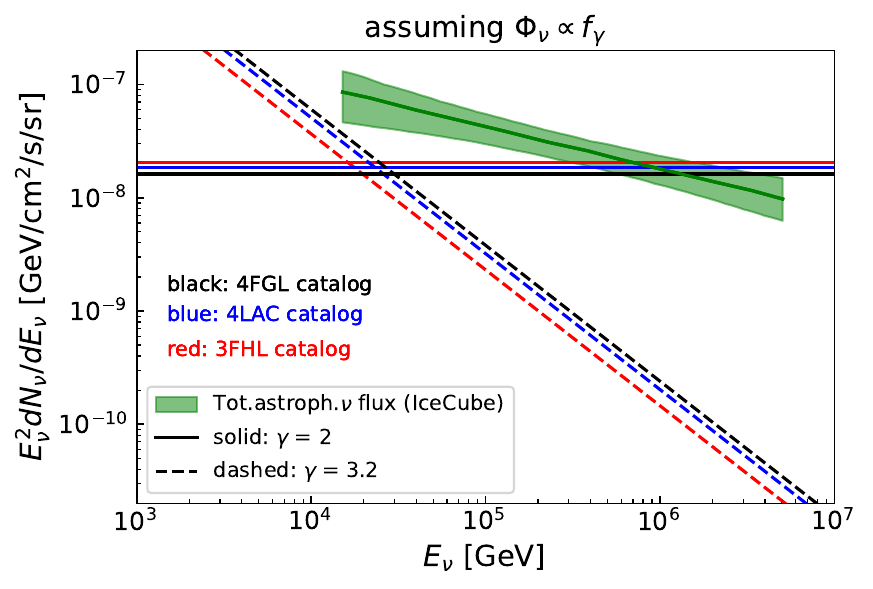}
\caption{{\it Left}: By requiring only 8 sources to exceed the discovery potential of IceCube point-source detection and assuming a proportionality for the neutrino flux, we can derive the constraints on the total neutrino flux contributed by the source catalogs. The left panel shows the case of 4LAC and $\gamma=2.0$ as a demonstration. Each blue point represents a 4LAC source and 8 points are above the {5$\sigma$ discovery potential} line. {\it Right}: Upper limits on the total neutrino flux from the sources in the 3FHL, {4FGL}, and 4LAC samples for two assumed spectral indices. The constraints are compared to the all-sky diffuse neutrino flux measured by IceCube \citep{IceCube:2021uhz} (green band). For both panels, we have assumed the proportionality $\phi_\nu\propto f_\gamma$ (see text for details).}

\label{fig:constraints}
\end{figure*}

Using the $5\sigma$ discovery potential lines in Ref.~\cite{icecube2022evidence} and assuming $\phi_\nu\propto f_\gamma$ \footnote{Since not all sources in the considered catalogs/samples have redshift information, here we only present the results for the $\phi_\nu\propto f_\gamma$ case. For the smaller populations of sources that have measured redshifts, the same analysis can also give the results for the $\phi_\nu\propto 1/D_L^2$ case.}, we estimate the maximum fractions that the gamma-ray samples can contribute to the total IceCube diffuse neutrino flux, which are shown in the right panel of Fig.~\ref{fig:constraints}. We find the constraints are weaker than those in L22. However, we believe this is a novel approach to derive a source population's contribution to the diffuse neutrino flux.

\subsection{Correlation study with 5BZCAT sources}
\label{sec:bzcat}
\begin{figure*}
\centering
\includegraphics[width=0.46\textwidth]{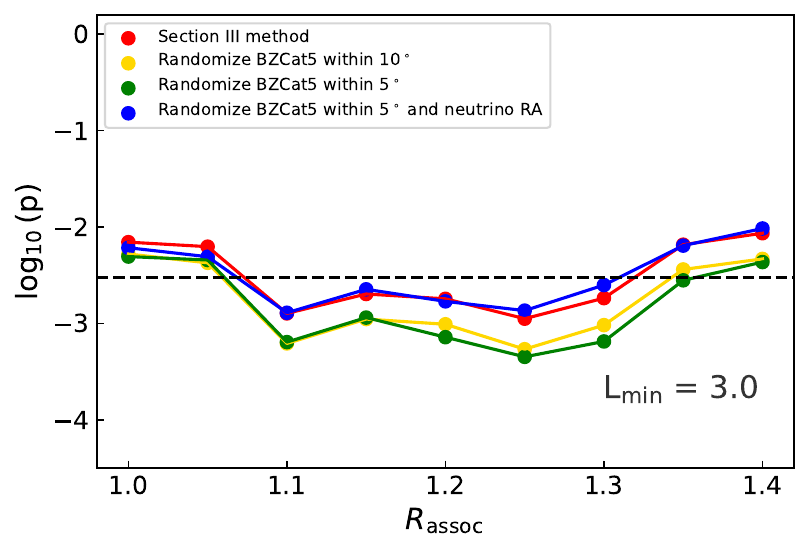}
\includegraphics[width=0.52\textwidth]{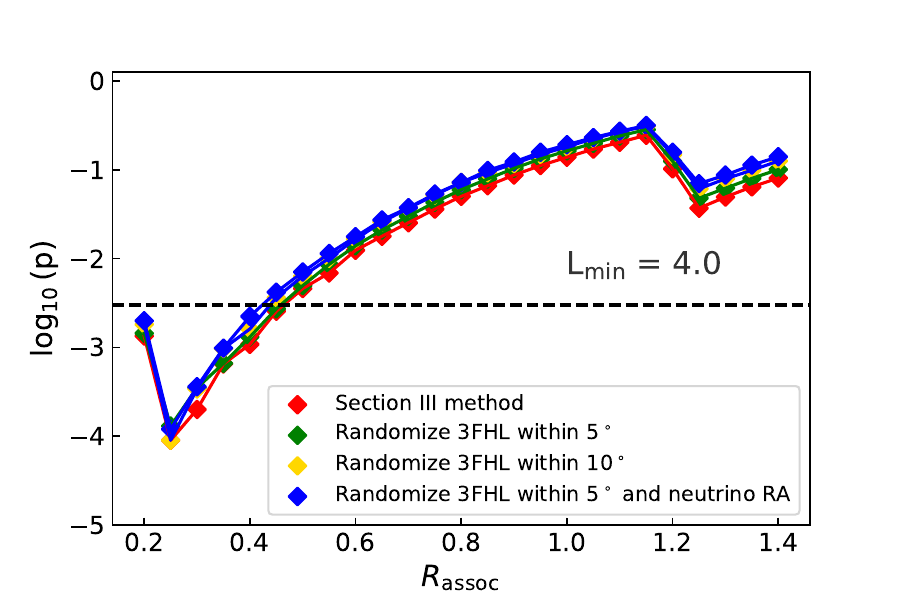}
\caption{These plots investigate how different ways of generating mock sources in the simulation impact on the derived p-values for the 5BZCAT sample (left panel) and the 3FHL sample (right panel).}
\label{fig:5bzcat}
\end{figure*}

This work adopts a similar analysis method of Ref.~\cite{Buson:2023irp}, which mainly studied the correlation between neutrino hotspots and blazars of the 5BZCAT catalog and claims a significant spatial correlation between the two (post-trial $2.47\sigma$ for the northern-sky analysis and $5.02\sigma$ if combining the northern and southern results).
To test our analysis pipeline and to verify the robustness of the claimed 5BZCAT-neutrino correlation, we also perform analysis using the 5BZCAT catalog.
The analysis procedure is the same as the above analysis of Fermi-LAT gamma-ray samples, only changing the source list to 5BZCAT.
The red line in Fig.~\ref{fig:5bzcat} and the last row in Table~\ref{tab:tab1} demonstrate the results of the 5BZCAT-neutrino correlation analysis.

As we can see, the $p_{\rm local}=1\times10^{-3}$ obtained in our analysis is larger than the one ($p_{\rm local}=5\times10^{-4}$) reported in Ref.~\cite{Buson:2023irp}.
We discuss the possible reason causing the difference.
The only point throughout our analysis that differs from Ref.~\cite{Buson:2023irp} is that in the simulation of deriving the p-value, we assign positions to mock sources in different ways.
Their Monte Carlo catalogs are generated by randomly shifting the sky position of the catalog sources within $10^\circ$ from their original position.
While we independently sample $l'$ and $b'$ from the $l$ list and $b$ list of the real sources, and the $(l',b')$ are shifted within $5^\circ$.
We note that the sampling approach employed in Sec.~\ref{sec_ana} requires an approximately symmetric source distribution in the Galactic coordinate system, which is appropriate for Fermi-LAT samples. However, as 5BZcat is a compiled catalog, the distribution of sources across the sky is apparently non-uniform, applying the Sec.~\ref{sec_ana} method may introduce methodological applicability issuses (but also note that we have used the KS-tests to ensure that the distribution matches the real catalogs).
If using exactly the same approach of generating mock sources as in Ref.~\cite{Buson:2023irp}, we get the result as shown as the yellow line in Fig.~\ref{fig:5bzcat}, which now coincides with their result.

To further investigate how large the manner of generating mock sources would affect the obtained p-values, we test some other methods of yielding the mock sources in the simulation.
They are: (1) randomize the 5BZCAT sources within $5^\circ$ radius from their original positions; 
(2) randomly shift the 5BZCAT sources within $5^\circ$ radius and meanwhile randomize the RA of the hotspots.
The corresponding results are depicted in Fig.~\ref{fig:5bzcat}, represented by green, and blue lines, respectively.

As shown in the figure, employing different simulation strategies yields minimum p-values in the range of $(0.4-1.3)\times10^{-3}$, corresponding to significances between $3.5\sigma$ and $3.2\sigma$. 
Taking into account a total of $3\times9 = 27$ trials\footnote{Note however that such a choice of the prior range for $R_{\rm assoc}$ may be controversial. Please see the argument in Ref.~\cite{Bellenghi:2023yza}.}, the post-trial p-values (significances) will be $(1.1-6.0)\times10^{-2}$ ($2.5\,\sigma$ to $2.1\,\sigma$).
Therefore, the analysis in this section demonstrates the way how mock sources are generated in the simulation has some impact on the results, but 
the pre-trial significance is $\gtrsim3\sigma$ for all the tested randomization approaches. The possible 5BZCAT-neutrino correlation is interesting and worthy of further investigation.

To also check whether the main results of Sec.~\ref{sec_results} are significantly affected by the randomization manners, we perform similar tests for the 3FHL sample and $L_{\rm min}=4.0$ case (which present the most significant correlation in our study). The results are displayed in the right panel of Fig.~\ref{fig:5bzcat}. It can be seen that the 3FHL/$L_{\rm min}=4.0$ result seems less affected, especially at the minimum p-value point. This is likely due to the smaller association radius.

\section{Summary}
\label{summary}
Currently, the observational connection between astrophysical sources' gamma-ray and neutrino emissions is still obscure. For instance, the neutrino flare\footnote{ Also note that the analysis by the IceCube collaboration with re-calibrated data yields a 2.7~$\sigma$ statistical significance for the flare \cite{IceCube:2021xar}, compared to 3.5~$\sigma$ reported from the former analysis.} of TXS 0506+056 occurred during a low gamma-ray state \cite{Padovani:2018acg,Fermi-LAT:2019hte}, and another candidate neutrino source NGC1068 exhibits relatively low gamma-ray flux in the Fermi-LAT observations \cite{Ajello:2023hkh,Ji:2023zit}. 
Although the proton-proton and proton-$\gamma$ processes that produce neutrinos will simultaneously accompany the production of gamma rays, {the gamma rays produced through hadronic processes could be reprocessed and emerge at lower energies \cite{Murase:2020lnu}.} However, considering that the GeV-TeV gamma rays observed by Fermi-LAT are the messengers with energies closest to IceCube's TeV-PeV neutrinos, gamma-ray emission and neutrino emission may be indirectly correlated. For instance, gamma-ray emission serves as a good indicator of a source’s ability to accelerate high-energy particles. The correlation between neutrino and gamma-ray observations is still worth investigating.
Hence, this work investigates the correlation between Fermi-LAT gamma-ray catalogs/samples and IceCube neutrino observations and is a follow-up to our last research \cite{Li:2022vsb}. We conduct spatial correlation analysis between various gamma-ray samples (4FGL \cite{Fermi-LAT:2019yla}, 4LAC \cite{Fermi-LAT:2019pir}, 3FHL \cite{Fermi-LAT:2017sxy}, etc.) and the hotspots in the IceCube p-value sky map of neutrino point-source scan (taken from \cite{icecube2022evidence}). By examining whether the number of source-hotspot associations significantly deviates from the expectation of complete chance coincidence, we probe the correlation between gamma-ray source samples and neutrino hotspots. 
Our results reveal weak evidence of a correlation between the 3FHL gamma-ray sample and neutrino hotspots, with a confidence level of $\sim1.7\sigma$.
The significance given by other samples is lower and also not significant, especially after considering the correction of trial factors. By examining the associated sources in 3FHL, we find that the $1.7\sigma$ correlation is mainly contributed by three already known neutrino sources/source candidates: TXS 0506+056, NGC 1068, and PKS 1424+240. That is, we do not discover any new potential neutrino source of neutrino emission. Furthermore, NGC 1068 (a Seyfert galaxy) belongs to a distinct AGN category compared to the two blazars. While the corona region surrounding its central supermassive black hole, hypothesized as the potential neutrino emission site, can produce gamma rays detectable in the Fermi-LAT band, the $>$10 GeV (3FHL) gamma-ray component is predominantly attributed to star-forming activity in the host galaxy \citep{Inoue:2021tcn,Blanco:2023dfp,Murase:2023ccp,Ajello:2023hkh} and appears to be not directly correlated with the observed neutrino emission.

\begin{acknowledgments}
This work is supported by the National Key Research and Development Program of China (Grant No. 2022YFF0503304), the National Natural Science Foundation of China (12373042, U1938201), the Programme of Bagui Scholars Programme (WXG) and Innovation
Project of Guangxi Graduate Education (YCBZ2024060).
\end{acknowledgments}

\bibliographystyle{apsrev4-1-lyf}
\bibliography{refer.bib}
\clearpage

\widetext
\begin{appendix}

\section{Neutrino hotspot list}
\label{app:hotspot}
In Table~\ref{tab:hotspot} we list all the hotspots with $L>3.0$ in the 2011-2020 IceCube skymap. The list is consistent with those reported in \cite{Buson:2023irp}. The only difference is that they report 82 total hotspots with $L>3.0$ while we here report 81. The reason is that there is a spot (RA=68.20, Dec=40.42, $L=3.07$), we call it A, that is only 1.55$^\circ$ away from another hotspot B (RA=68.20, Dec=38.87, $L=3.22$). Within the 1.5-degree radius around A, there are other pixels with p-values greater than A (these pixels are not considered new hotspots because they are less than 1.5$^\circ$ away from B). These signs suggest that A might be an extension of hotspot B rather than a new hotspot. Therefore, we do not consider it as an independent hotspot. Since this hotspot lies within the Galactic plane with $|b| < 10^\circ$, whether or not it is included does not affect the results of our analysis.
\begin{table}[!htbp]
\centering
\caption{List of neutrino hotspots with $L>3.0$ in the 2011-2020 IceCube skymap, {which is consistent with those reported in \citep{Buson:2023irp}}.}
\begin{tabular}{p{1.5cm}p{1.5cm}p{1.5cm}p{1.5cm}p{1.5cm}p{1.0cm}}
\hline
\hline
\text{RA $[^\circ]$} & \text{Dec $[^\circ]$} & $L$ & \text{RA $[^{\circ}]$} & \text{Dec $[^{\circ}]$} & $L$  \\
	\hline
 	16.5 & 73.5 & 3.01 & 26.72 & 0.45 & 3.02 \\
	47.64 & 2.09 & 3.03 & 283.87 & 65.51 & 3.04 \\
	208.88 & 77.73 & 3.05 & 122.7 & 22.19 & 3.06 \\
	215.86 & 2.54 & 3.06 & 241.52 & 38.49 & 3.07 \\
	233.79 & 38.11 & 3.07 & 258.05 & 32.62 & 3.09 \\
	39.02 & 66.44 & 3.09 & 284.24 & 23.64 & 3.09 \\
	35.51 & 24.13 & 3.11 & 26.89 & 20.74 & 3.12 \\
	288.46 & 15.4 & 3.12 & 321.86 & 12.64 & 3.13 \\
	31.46 & 34.59 & 3.13 & 265.45 & 43.61 & 3.13 \\
	99.49 & 10.66 & 3.14 & 45.18 & 31.74 & 3.14 \\
	64.86 & 2.69 & 3.15 & 149.94 & 37.17 & 3.17 \\
	23.73 & 31.04 & 3.17 & 161.19 & 27.28 & 3.17 \\
	230.45 & 23.32 & 3.21 & 339.71 & 48.14 & 3.21 \\
	342.33 & 44.6 & 3.21 & 307.56 & 43.41 & 3.22 \\
	68.2 & 38.87 & 3.22 & 306.74 & 19.16 & 3.22 \\
	86.54 & 58.73 & 3.23 & 53.26 & 51.45 & 3.25 \\
	340.31 & 7.33 & 3.27 & 237.66 & 18.68 & 3.29 \\
	171.57 & 42.81 & 3.33 & 292.85 & 1.64 & 3.33 \\
	49.22 & 6.43 & 3.33 & 114.43 & 28.29 & 3.35 \\
	174.73 & 18.37 & 3.38 & 250.93 & 50.87 & 3.4 \\
	351.56 & 13.09 & 3.4 & 59.77 & 5.83 & 3.41 \\
	171.21 & 26.44 & 3.44 & 215.16 & 13.09 & 3.44 \\
	306.11 & 45.39 & 3.47 & 230.74 & 45.98 & 3.49 \\
	309.9 & -0.6 & 3.49 & 292.5 & 33.33 & 3.53 \\
	294.5 & 73.5 & 3.55 & 36.74 & 15.4 & 3.6 \\
	144.49 & 9.44 & 3.6 & 302.87 & 1.79 & 3.61 \\
	201.86 & 50.48 & 3.64 & 191.07 & 40.62 & 3.64 \\
	7.91 & 15.4 & 3.67 & 78.12 & 57.02 & 3.67 \\
	208.3 & 26.28 & 3.69 & 215.4 & 76.26 & 3.72 \\
	244.86 & 38.68 & 3.73 & 20.92 & 25.61 & 3.77 \\
	245.21 & 16.02 & 3.8 & 40.43 & -2.24 & 3.83 \\
	170.16 & 27.78 & 3.83 & 182.46 & 39.64 & 3.92 \\
	121.33 & 50.09 & 4.0 & 202.68 & 33.87 & 4.01 \\
	322.12 & 67.93 & 4.03 & 110.21 & 11.42 & 4.04 \\
	82.44 & 32.8 & 4.05 & 77.34 & 5.53 & 4.13 \\
	216.91 & 23.81 & 4.18 & 299.53 & 29.83 & 4.19 \\
	179.33 & 52.42 & 4.22 & 105.82 & 1.04 & 4.32 \\
	177.89 & 23.16 & 4.32 & 9.67 & 7.48 & 4.34 \\
	208.12 & 23.16 & 4.58 & 180.18 & 42.21 & 4.79 \\
	76.29 & 12.79 & 5.19 & 297.42 & 27.45 & 5.28 \\
	40.78 & 0.15 & 6.75 \\
\hline
\hline
\end{tabular}
\label{tab:hotspot}
\end{table}

\section{Tests of the effects of the randomization approach and randomization radius on results}
\label{app:cross-check}

\begin{figure*}[h]
\centering
\includegraphics[width=0.45\textwidth]{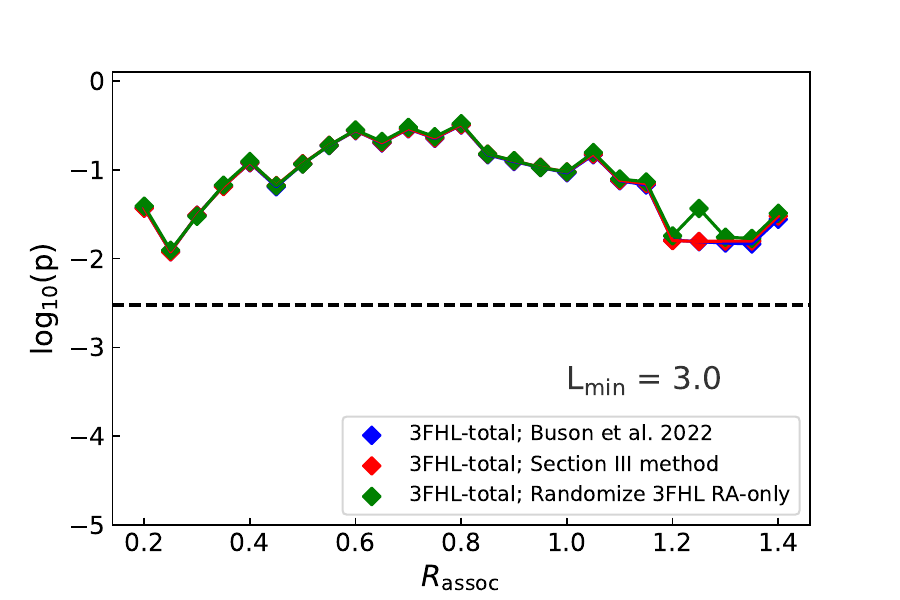}
\includegraphics[width=0.45\textwidth]{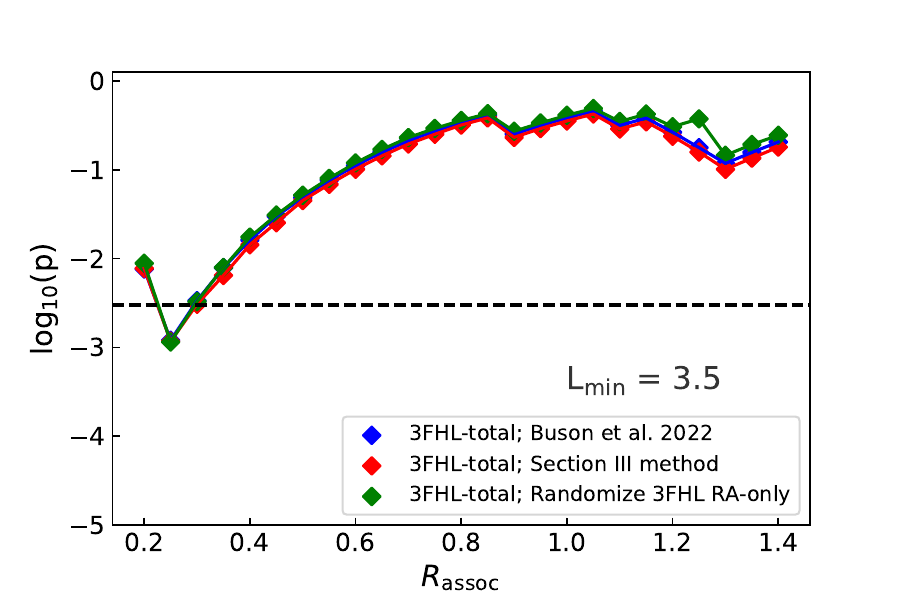}
\includegraphics[width=0.45\textwidth]{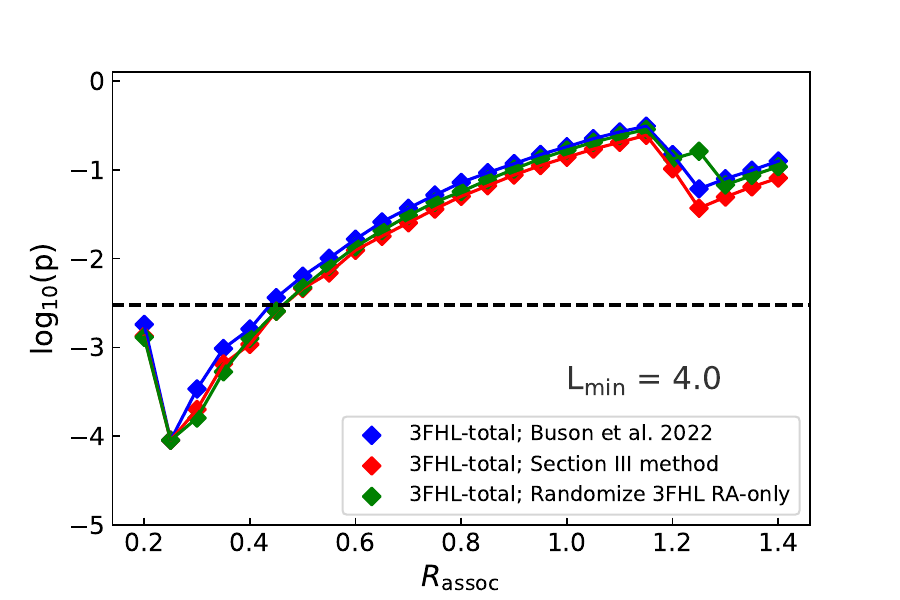}
\caption{Correlation results obtained using 3 different randomization methods for the 3FHL TOTAL sample and different choices of $L_{\rm min}$.}
\label{fig:cross-check}
\end{figure*}

The randomization approaches used to generate mock sources may affect the obtained results of our work. Here we present the results derived by adopting other randomization approaches as a cross-check of our results in the main text. We employ the 3FHL TOTAL sample for the tests. We consider two other ways of the randomization, 1) only randomizing the sources’ right ascension (i.e., the one used in \citet{Plavin:2020mkf,Buson:2023irp}), 2) randomizing the real sources within a $10^\circ$ radius from their original positions (the one used in \citet{Buson:2023irp}). It can be seen in Fig.~\ref{fig:cross-check} that the results obtained by the different sampling methods have small differences in the quantitative $\log_{10}(p)$ values, but do not affect the main results of this paper.

\begin{figure*}[h]
\centering
\includegraphics[width=0.55\textwidth]{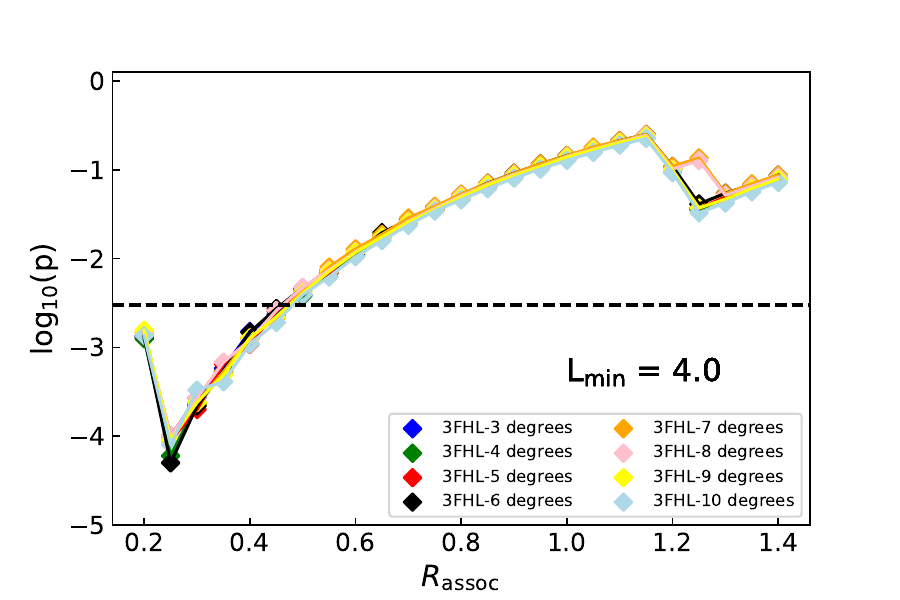}
\caption{Correlation results obtained adopting different randomization radii for the 3FHL TOTAL sample and $L_{\rm min}=4.0$.}
\label{fig:different-deg}
\end{figure*}

We also investigate the impact of varying the randomization radius on the results, which is shown in Fig.~\ref{fig:different-deg}. We choose the case of the 3FHL TOTAL sample and $L>4.0$ (which shows the most significant correlation in our results) for demonstration. 
As shown in Fig.~\ref{fig:different-deg}, no significant difference is observed in this test for different randomization radii, indicating that the 5$^\circ$ radius is an appropriate choice for our analysis of Fermi-LAT catalogs.

\end{appendix}

\end{document}